\documentclass[preprint,12pt]{elsarticle}
\usepackage{amssymb}
\usepackage{url}
\usepackage{algorithm}
\usepackage{algpseudocode}
\usepackage{lscape}
\usepackage{color}
\usepackage{amsmath}
\usepackage{amsthm}
\newtheorem{theorem}{Theorem}
\newtheorem{corollary}{Corollary}
\setcitestyle{authoryear,open={(},close={)}}
\usepackage{lineno}
\makeatletter
\def\ps@pprintTitle{%
 \let\@oddhead\@empty
 \let\@evenhead\@empty
 \def\@oddfoot{\reset@font\hfil\thepage\hfil}
 \let\@evenfoot\@oddfoot
}

\makeatletter

\begin{document}
\begin{frontmatter}

\title{Nonparametric goodness of fit tests for Pareto type-I distribution with complete and censored data}

\author[label1]{Avhad Ganesh Vishnu}
\ead{avhadgv@gmail.com}
\author[label1]{Ananya Lahiri}
\author[label2]{Sudheesh K. Kattumannil}
\address[label1]{Department of Mathematics and Statistics, Indian Institute of Technology, Tirupati, India;}
\address[label2]{Statistical Sciences Division, Indian Statistical Institute, Chennai, Tamil Nadu, India}

\begin{abstract}
Two new goodness of fit tests for the Pareto type-I distribution for complete
and right censored data are proposed using fixed point characterization based
on Stein’s type identity. The asymptotic distributions of the test statistics under both the null and alternative hypotheses are obtained. The performance of the proposed tests is evaluated and compared with existing tests through a Monte Carlo simulation experiment. The newly proposed tests exhibit greater power than existing tests for the Pareto type-I distribution. Finally, the methodology is applied to real-world data sets.
\end{abstract}

\begin{keyword}
 Goodness of fit testing $\cdot$ Stein's identity $\cdot$ Pareto distribution $\cdot$ U-statistics $\cdot$ Censored data.
\end{keyword}
\end{frontmatter}

\section{Introduction}\label{sec1}
The Pareto distribution has been of significant interest in various sectors due to its extensive applicability and significance in modeling events that exhibit heavy-tailed distributions. Because of its widespread usage, considerable interest has been attracted from researchers, leading to the development of various versions such as type-I, II, III, IV, and generalized Pareto distributions. A comprehensive discussion of these multiple types of Pareto distributions, elucidating the relationships between them, is provided by \cite{Arnold}. Several tests have been designed to assess the notion that observed data follows a Pareto distribution because many forms of Pareto distributions have found widespread application. This paper considers the goodness of fit test problem for the Pareto type-I distribution.

Confirming data alignment with a particular family of distributions is crucial in data analysis, and various goodness-of-fit tests serve this purpose effectively. Characterizing a specific family of distributions is a defining property unique to that family. For further insights into characterizations, refer to \cite{galambos2006characterizations}. Characterizations effectively differentiate one distributional family from others, making them valuable for goodness-of-fit testing purposes. 

Goodness-of-fit tests tailored for the Pareto type-I distribution have been extensively studied in the literature. \cite{chu2019review} and \cite{ndwandwe2023testing}  present a comprehensive review of tests designed to evaluate the fit of data to the Pareto distribution, focusing specifically on the Pareto type-I distribution. Recently, \cite{ngatchou2024classes} proposed new classes of tests based on a characterization of the Pareto distribution involving order statistics. Tests based on different characterizations for the Pareto distribution have been approached by several authors, including \cite{obradovic2015goodness}, \cite{volkova2016goodness}, \cite{bojana2016}, \cite{akbari2020characterization} and \cite{allison2022distribution} among others.

It is important to note that all the goodness-of-fit tests discussed have been developed for complete data sets. However, censoring, particularly right-censoring, is common in lifetime and survival analysis. Therefore, goodness-of-fit tests for the Pareto type-I distribution accommodating right-censored observations must be developed, ensuring their applicability to real-world data where censoring occurs.

A moment identity for a random variable whose distribution belongs to the exponential family was introduced by \cite{stein1972bound}. This identity, known as Stein's type identity, has been extensively investigated in the statistical literature due to its importance in inference procedures. Comprehensive discussions on Stein's type identity applicable to a wide range of probability distributions and their associated characterizations can be found in the works of \cite{kattumannil2009stein}, \cite{kattumannil2012moment}, \cite{kattumannil2016generalized}, and \cite{anastasiou2023stein}, among others. Using Stein's type identity, a fixed point characterization for univariate distributions was established by \cite{betsch2021fixed}. Motivated by this, U-statistic-based goodness of fit tests for the Pareto type-I distribution for complete and right-censored data are developed.

The paper is organised as follows. Section \ref{sec2} presents a characterization of the Pareto type-I distribution, followed by an introduction to a new class of tests designed for this distribution. Two test statistics are proposed, and their asymptotic distribution for complete and censored observations is obtained in Section \ref{sec3} and Section \ref{sec4}. Moving on to Section \ref{sec5}, the finite-sample performance of the newly proposed tests is assessed through Monte Carlo simulations and compared with other existing tests. In Section \ref{sec6}, all tests are applied to real-world data sets. Finally, the paper is concluded, and a summary is provided in Section \ref{sec7}.

\section{New characterization of the Pareto type-I distribution}\label{sec2}

Using Stein's type identity, \cite{betsch2021fixed} developed a fixed point characterization for a large class of absolutely continuous univariate distributions. The Stein characterization for semi-bounded support states that a real-valued random variable $X$ has density $f$ supported by $[L, \infty)$ and holds the following conditions:
\begin{enumerate}
    \item $P\big(X\in [L,\infty) \big)=1$,\\
    \item $E\bigg[\Big|\dfrac{f'(X)}{f(X)} \Big|\bigg]<\infty$,~ and \\
    \item $E\bigg[\Big|\dfrac{Xf'(X)}{f(X)} \Big|\bigg]<\infty$.
\end{enumerate}
if, and only if, the distribution function of $X$ has the form 
\begin{equation*}
   F(t)= E \bigg[-\dfrac{f'(X)}{f(X)} \Big(\min (X,t) -L \Big) \bigg], ~~~ t>L. 
\end{equation*}
One can refer to supplementary material of \cite{betsch2021fixed} for proof.  

Let $P(\alpha)$ be the class of Pareto type-I distribution with the distribution function 
\begin{equation}\label{eq1}
    F(x) = 1- x^{-\alpha}, ~~ x\geq 1, ~~ \alpha>0, 
\end{equation} 
and density function 
\begin{equation*}
    f(x) = \alpha x^{-(\alpha+1)}, ~~ x\geq 1, ~~ \alpha>0, 
\end{equation*} 
where $\alpha$ is the shape parameter. 
    
Using \cite{betsch2021fixed}, Theorem 3, we utilize the following fixed point characterization based on Stein’s type identity for the Pareto type-I distribution to develop the test.
\begin{theorem}\label{thm1}
  Let $X$ be a positive random variable with $E(X)<\infty$.  Then $X$ has the Pareto type-I distribution with shape parameter $\alpha$ if, and only if, the distribution function of $X$ has the form 
    \begin{equation*}
        F(t)= E\bigg[\dfrac{(\alpha+1)}{X} \big(\min(X,t)-1\big) \bigg],  ~~t>1.
    \end{equation*}
\end{theorem}  
Based on random samples $X_1, X_2, \ldots, X_n$ drown from the distribution function $F$ with the support $\mathcal{X}(= \mathcal{R})$. Also a measurable function $h: \mathcal{X}^k \to \mathcal{R}$, refer to as a symmetric kernel of degree $k(\leq n)$, the null hypothesis is tested as
\begin{align*}
    H_0 &: F \in P(\alpha),
\end{align*}
against the alternative
\begin{align*}
    H_1 &: F \notin P(\alpha).
\end{align*}
 
Using Theorem \ref{thm1}, we introduced two test statistics as integral type statistic ($\Delta_{I}$) and Cramér–von Mises type statistic ($\Delta_{M}$) as
\begin{align}
  \Delta_{I}  &= \int_1^\infty \Big( E\Big[\dfrac{(\alpha +1)}{X} \big(\min(X,t)-1\big) \Big]-F(t) \Big) dF(t),
\end{align}
 and
\begin{align}
   \Delta_{M} &=\int_1^\infty \Big( E\Big[\dfrac{(\alpha+1)}{X} \big(\min(X,t)-1\big) \Big]-F(t) \Big)^2 dF(t).
\end{align}

\section{$ \Delta_{I}$ : Test statistics}\label{sec3}

In this section, the integral-type statistic is discussed. 
The following subsection examines the properties of the test statistic for complete observations.
\subsection{ Uncensored case}\label{sec3.1}
To develop the test, a departure measure is defined that discriminates between the null and alternative hypotheses. Consider $ \Delta_{I}$ given by
\begin{equation}\label{eqT1}
    \Delta_{I} = \int_{1}^{\infty} \Big(E\Big[\dfrac{(\alpha+1)}{X}\big(\min(X,t)-1 \big)\Big]- F(t)  \Big) dF(t).
\end{equation}

In the context of Theorem \ref{thm1}, $\Delta_{I}$ is non-zero under the alternative hypothesis $H_1$ and zero under the null hypothesis $H_0$. 
Since we are using the U-statistics theory to find the test statistic, we represent $ \Delta_{I}$ as an expectation of a function of the random variables. Consider
\begin{small}
\begin{align}\label{eq2}
    \Delta_{I} &= \int_{1}^{\infty}E\Big[\dfrac{\alpha+1}{X}\big(\min(X,t)-1 \big)\Big]dF(t) - \int_{1}^{\infty} F(t) dF(t) \nonumber\\
    &= \int_{1}^{\infty} \int_{1}^{\infty}\Big(\dfrac{\alpha+1}{x}\Big)\min(x,t) dF(x)dF(t)- \int_{1}^{\infty} \int_{x=1}^{\infty}\Big(\dfrac{\alpha+1}{x}\Big) dF(x)dF(t) - \dfrac{1}{2}\nonumber\\
    &= \int_{1}^{\infty} \int_{1}^{\infty}\Big(\dfrac{\alpha+1}{x}\Big)(xI(x<t)) dF(x)dF(t) \nonumber\\
    &\hspace{0.7cm} + \int_{1}^{\infty} \int_{1}^{\infty}\Big(\dfrac{\alpha+1}{x}\Big) (t I(t<x)) dF(x)dF(t) \nonumber \\
    &\hspace{0.7cm}- \int_{1}^{\infty} \int_{1}^{\infty}\Big(\dfrac{\alpha+1}{x}\Big) dF(x)dF(t) - \dfrac{1}{2}\nonumber\\
     &=(\alpha+1) \bigg( \int_{1}^{\infty} \int_{1}^{t} dF(x)dF(t) + \int_{1}^{\infty} \int_{1}^{x} \dfrac{t}{x} dF(x)dF(t)\nonumber\\
     &\hspace{2.2cm} - \int_{1}^{\infty} \int_{1}^{\infty} \dfrac{1}{x}  dF(x)dF(t)\bigg) - \dfrac{1}{2}\nonumber\\
     &= (\alpha+1) E\Big[ I(X_1< X_2) + \dfrac{X_2}{X_1}I(X_2<X_1)-\dfrac{1}{X_1}\Big] -\dfrac{1}{2}\nonumber\\
     &= (\alpha+1) E\Big[\dfrac{X_2}{X_1}I(X_2<X_1)-\dfrac{1}{X_1} \Big] +\dfrac{\alpha}{2},
\end{align}
\end{small}
where $I(A)$ denotes the indicator function of a set $A$. The symmetric kernel $h_1$ is defined as
\begin{align}\label{h(1)}
    h_1(X_1, X_2)&= \dfrac{1}{2} \Big[\dfrac{X_2}{X_1}I(X_2<X_1)+  \dfrac{X_1}{X_2} I(X_1<X_2)-\dfrac{1}{X_1} -\dfrac{1}{X_2}  \Big]. 
\end{align}
Then  a U-statistic defined by
\begin{equation*}
    U = \dfrac{2}{n(n-1)} \sum_{i=1  }^{n} \sum_{j < i, j=1}^{n} h_1(X_i, X_j), 
\end{equation*}
is an unbiased estimator of $E\Big(\dfrac{X_2}{X_1}I(X_2<X_1)-\dfrac{1}{X_1}\Big)$. 
A consistent moment-based estimator of $\alpha$, under the condition $\alpha>1$ for the Pareto type-I distribution (see \cite{quandt1964old}) is obtained by 
\begin{equation*}
    \widehat{\alpha} =\dfrac{\Bar{X}}{(\Bar{X}-1)},
\end{equation*}
where $\Bar{X}$ denotes the sample mean of the random variable $X$.
Hence, the test statistic is given by 
\begin{equation}\label{eq3}
    \widehat{\Delta}_{I}= (\widehat{\alpha}+1)U +\dfrac{\widehat{\alpha}}{2}.
\end{equation}
Note that ${\Delta}_{I}$ be the U-statistic with symmetry kernel $h$ is defined as, 
\begin{small}
\begin{align}\label{h()}
    h(X_1, X_2)&= \dfrac{(\alpha+1)}{2} \Big[\dfrac{X_2}{X_1}I(X_2<X_1)+  \dfrac{X_1}{X_2} I(X_1<X_2)-\dfrac{1}{X_1} -\dfrac{1}{X_2} \Big]+\dfrac{\alpha}{2}, 
\end{align}
\end{small}
such that $E(h(X_1, X_2))= {\Delta}_{I}$.

The null hypothesis $H_0$ is rejected in favor of the alternative hypothesis $H_1$ for a large value of $|\widehat{\Delta}_{I}|$.

Next, the asymptotic properties of the test statistic are studied. Since $\widehat{\alpha}$ and $U$ are U-statistics, they are consistent estimators of $\alpha$ and $E\Big(\dfrac{X_2}{X_1}I(X_2<X_1)-\dfrac{1}{X_1}\Big)$, respectively (see \cite{lehmann1951consistency}). Hence, the following result is straightforward.

\begin{theorem}\label{thm2}
    As $\widehat{\alpha}$ be the consistent estimator of $\alpha$. Under $H_1$, as $n \to \infty$, $\widehat{\Delta}_{I}$ converges in probability to $\Delta_{I}$. 
\end{theorem}

\begin{theorem}\label{thm3}
   As $n \to \infty$, $\sqrt{n}(\widehat{\Delta}_{I}- \Delta_{I})$ converges in distribution to a normal random variable with mean zero and variance $4\sigma^2$, where $\sigma^2$ is obtained by 
\begin{equation}\label{sigma^2}
    \sigma^2= Var[E(h^{}(X_1, X_2)|X_1)]. 
\end{equation}
\begin{proof}
     Define 
     \begin{equation*}
         \widetilde{\Delta}_{I}  = (\alpha+1)U+\dfrac{\alpha}{2}.
     \end{equation*}
Consider
\begin{align*}
    \sqrt{n}(\widehat{\Delta}_{I}- \Delta_{I})&= \sqrt{n}(\widehat{\Delta}_{I}- \widetilde{\Delta}_{I})+\sqrt{n}(\widetilde{\Delta}_{I}- \Delta_{I}). 
\end{align*}
The first term as
\begin{align*}
    \sqrt{n}(\widehat{\Delta}_{I}- \widetilde{\Delta}_{I})&=\sqrt{n} \Big((\hat{\alpha}+1)U + \frac{\hat{\alpha}}{2}-(\alpha+1)U - \frac{{\alpha}}{2})\Big)\\
    &= \sqrt{n}\Big((\hat{\alpha}-\alpha)U + \Big(\frac{\hat{\alpha}}{2} - \frac{{\alpha}}{2}\Big)\Big).
\end{align*}
Since $\widehat{\alpha}$ be the consistent estimator of $\alpha$. We know
\begin{equation*}
    \widehat{\alpha}\xrightarrow{P} \alpha, ~ U\xrightarrow{P} E(U) \implies \widehat{\alpha} U \xrightarrow{P} \alpha E(U) \implies (\widehat{\alpha}-\alpha)U \xrightarrow{P} 0.
\end{equation*}
Using Chebyshev's inequality, 
\begin{align*}
   \sqrt{n}\Big((\hat{\alpha}-\alpha)U + \Big(\frac{\hat{\alpha}}{2} - \frac{{\alpha}}{2}\Big)\Big)\xrightarrow{P} 0 \implies 
   \sqrt{n}(\widehat{\Delta}_{I}- \widetilde{\Delta}_{I}) \xrightarrow{P} 0.
\end{align*}
Also note that $E(\widetilde{\Delta}_{I})=\Delta_{I}$.
This leads to  $\sqrt{n}(\widetilde{\Delta}_{I}- \Delta_{I}) =   \sqrt{n}(\widetilde{\Delta}_{I}-E( \widetilde{\Delta}_{I}))$.
 Now, we observe that $\widetilde{\Delta}_{I}$ is a U-statistic with the symmetric kernel  $h^{}(X_1, X_2)$.  
 Using the central limit theorem for U-statistics, the asymptotic normality of $\widetilde{\Delta}_{I}$ is established (see \cite{lee2019u}, Theorem 1, page 76). The asymptotic variance is $4\sigma^2$, where $\sigma^2$ is given by  
\begin{equation*}
    \sigma^2= Var[E(h^{}(X_1, X_2)|X_1)]. 
\end{equation*}
Note that
 \begin{align*}
   &E(h^{}(X_1, X_2)|X_1=x)\\
   &= \dfrac{(\alpha+1)}{2}  E\bigg[ \dfrac{X_2}{x} I(X_2<x)+ \dfrac{x}{X_2}I(x<X_2) -\dfrac{1}{x} -\dfrac{1}{X_2} \bigg] \\
     &= \dfrac{(\alpha+1)}{2} \bigg[  \int_{1}^{x} \dfrac{y}{x} dF(y)+ \int_{x}^{\infty} \dfrac{x}{y} dF(y) -\dfrac{1}{x} - \int_{1}^\infty \dfrac{1}{y} dF(y)  \bigg]. 
 \end{align*}
So, the second term 
 $\sqrt{n}(\widetilde{\Delta}_{I}- \Delta_{I})\xrightarrow{d} N(0,4\sigma^2)$. 
 Now using Slutsky's theorem $\sqrt{n}(\widehat{\Delta}_{I}- \Delta_{I})\xrightarrow{d} N(0,4\sigma^2)$.
 
\end{proof}
\end{theorem}

 Under the null hypothesis $H_0$, $\Delta_{I}=0$. Hence, the following corollary is obtained.

\begin{corollary}\label{corl1}
    Under $H_0$, as $n \to \infty$, $\sqrt{n}\widehat{\Delta}_{I}$ converges in distribution to a normal random variable with mean zero and variance $4\sigma_{0}^2$, where $\sigma_{0}^2$ is obtained by evaluating (\ref{sigma^2}) under $H_0$.
\end{corollary}

The asymptotic critical region for the scale-invariant test can be obtained using Corollary \ref{corl1}. Let $\widehat{\sigma}_0^2$ be a consistent estimator of $\sigma_0^2$. The null hypothesis $H_0$ is rejected in favor of the alternative hypothesis $H_1$ at a significance level of $\gamma$ if
\begin{equation*}
    \dfrac{\sqrt{n}|\widehat{\Delta}_{I}|}{2\widehat{\sigma}_0}> Z_{\gamma/2}, 
\end{equation*}
where $Z_\gamma$ is the upper $\gamma$- percentile point of the standard normal distribution. Since it is difficult to find the null variance $\sigma_0^2$, we obtained the critical region of the test using the bootstrap procedure. The lower $(C_1)$ and upper $(C_2)$ critical points are identified in a such way that $P(\widehat{\Delta}_{I}<C_1)= P(\widehat{\Delta}_{I}>C_2)= \gamma/2$.

\subsection{ Right censored case }\label{sec3.2}
The suggested testing methodology is now extended to incorporate censored observations. Let $X$ represent the lifetime and $C$ censoring time with density function $g$ and distribution function $G$. The observed lifetime is $Y = \min(X, C)$ and $\delta = I(X \leq C)$ is the censoring indicator. Independence of lifetimes and censoring times is assumed. The test explained in Section \ref{sec3.1} is then modified using  $n$ independent and identically distributed random vectors {$(Y_i, \delta_i), 1 \leq i \leq n $} drawn from $(Y, \delta)$.

Define $R_i(t)= I(Y_i\geq t)$ as the counting process corresponding to censoring random variable for the $i$-th subject and $N_i^c(t)= I(Y_i \leq t, \delta_i=0)$ as the counting process of the censored variable where $\delta_i=0$. Furthermore, let $\alpha_c(t)= g(t)/\Bar{G}(t)$ be the hazard rate function of censoring variable $C$. Given this counting process $N_i^c(t)$, the martingale associated with it is given by (see \cite{andersen2012statistical})

\begin{equation*}
    M_i^c(t)= N_i^c(t)- \int_{0}^{t} R_i(u)\alpha_c(u) du, ~~ i=1,2,\ldots,n.
\end{equation*} 

An estimator of the survival function of censoring variable $C$ under right-censoring (see \cite{satten2001kaplan}) denoted by $\hat{S}_c(\cdot)$, is given by
\begin{equation*}
   \hat{S}_c(t)= \prod_{t_i\leq t}\Big(1- \dfrac{N^c(t_i)}{R(t_i)} \Big),
\end{equation*}
where $N^c(t)=\sum_{i=1}^n N^c_i(t)$ is the number of death events of the counting process corresponding to the censoring time and $R(t)= \sum_{i=1}^n R_i(t)$ is the number of subject at risk just prior to the time $t$ .

Since the right-censored data is analysed using U-statistics theory, the same departure measure $\Delta_{I}$ defined in (\ref{eqT1}) is used here as well. As the distribution of the censoring time is continuous, \( \hat{S}_c(t-) \) equals \( \hat{S}_c(t) \). A U-statistic for the right-censored data is given by the definition provided by \cite{datta2010inverse}, as follows:
\begin{equation*}
    \widehat{\Delta}_{I_c}^* = \dfrac{1}{n(n-1)}\sum_{i=1  }^{n} \sum_{j < i, j=1}^{n} \dfrac{\Big(  \dfrac{Y_j}{Y_i} I(Y_j<Y_i) +   \dfrac{Y_i}{Y_j} I(Y_i<Y_j)-\dfrac{1}{Y_i}-\dfrac{1}{Y_j}\Big)\delta_i \delta_j }{\widehat{S}_c(Y_i)\widehat{S}_c(Y_j)},
\end{equation*}
where $\dfrac{\delta_i \delta_j }{\widehat{S}_c(\cdot)\widehat{S}_c(\cdot)}$ is the weight function. 
Next, the moment-based estimator of $\alpha(>1)$ under right-censored data is obtained as
\begin{equation*}
   \Bar{X}_c = \dfrac{1}{n} \sum_{i=1  }^{n} \dfrac{Y_i \delta_i} {\widehat{S}_c(Y_i)} ~~ \text{and} ~~ \widehat{\alpha}_c = \dfrac{\Bar{X}_c}{(\Bar{X}_c-1)}, 
\end{equation*}
where $\Bar{X}_c$ denotes the censored sample mean of the random variable $X$. The moment estimator of $\alpha$ is
\begin{equation*}
  \hat{\alpha}=\dfrac{\Bar{X}}{\Bar{X}-1}.  
\end{equation*}
Using Theorem $3$ of \cite{kattumannil2021non}, it can be proven that $\Bar{X}_c$ is consistent estimator of $\Bar{X}$. It can also be easily verified that $\widehat{\alpha}_c$ is a consistent estimator of $\alpha$.
As a result, the test statistic is obtained as 
\begin{equation}
   \widehat{\Delta}_{I_c} = ( \widehat{\alpha}_c+1)\widehat{\Delta}_{I_c}^* +\frac{\widehat{\alpha}_c}{2} . 
\end{equation}

Let
\begin{align*}
    h_1(x)&= \dfrac{(\alpha+1)}{2} E\Big(\dfrac{Y_2}{x}I(Y_2< x)
    + \dfrac{x}{Y_2}I(x<Y_2)-\dfrac{1}{x}-\dfrac{1}{Y_2}\Big)+\dfrac{\alpha}{2},
\end{align*}
such that $E(h(Y_1, Y_2) | Y_1 = x) = h_1(x)$, where $h(\cdot, \cdot)$ is defined in (\ref{h()}). Now, the empirical sub-distribution function of the pair $(X_i, Y_i)$ is defined as
\begin{align*}
    H_c(x,t)&= P(X_1\leq x, Y_1\leq t, \delta=1),~~ x \in \mathcal{X}, ~t\geq 0,
 \end{align*}   
 $\Bar{K}(t)= E(R_1(t))$, and weight function as
 \begin{align*}
   w(t)&= \dfrac{1}{\Bar{K}(t)} \int_{\mathcal{X}\times [0, \infty)} \dfrac{h_1(x)}{S_c(y)} I(y>t)dH_c(x,y), ~~  t\geq 0,
\end{align*}
where $I(y>t)$ be the risk indicator. The proof of the following theorem can be done similarly to Theorem $3$ with a particular choice of the kernel function ( see \cite{datta2010inverse}).  

\begin{theorem}\label{thmcen}
Assume that
\begin{equation*}
    E\big[ h(Y_1, Y_2)  \big]^2 <\infty, ~ \int_{\mathcal{X}\times [0, \infty)}\dfrac{h^2_1(x)}{S^2_c(y)}dH_c(x,y) <\infty ~~ \text{and} ~~\int_0^\infty w^2(t) \alpha_c(t) dt<\infty.  
\end{equation*}
As \( n \to \infty \), the distribution of \( \sqrt{n}(\widehat{\Delta}_{I_c} - \Delta_{I}) \) converges to a normal distribution with a mean of zero and a variance \( 4\sigma^2_{c} \). The variance \( \sigma^2_{c} \) is obtained as follows:
\begin{equation*}
    \sigma^2_{c}= Var\bigg( \dfrac{h_1(X)\delta_1}{S_c(Y_1)}+\int_0^\infty w(t) dM_1^c(t) \bigg). 
\end{equation*}
\end{theorem}

As suggested by \cite{datta2010inverse}, the reweighted average technique is used to simplify the asymptotic analysis. Therefore, the reweighted approach is used to find an estimator of $\sigma^2_{c}$. An estimator of $\sigma^2_{c}$ is given by
\begin{equation*}
    \widehat{\sigma}^2_{c}= \dfrac{4}{(n-1)} \sum_{i=1}^n (V_i- \Bar{V})^2,  
\end{equation*}
where
\begin{align*}
    \widehat{h}_1(X)&=  \dfrac{1}{n}\sum_{j=1}^n \dfrac{h(X,Y_j)\delta_j}{\widehat{S}_c(Y_j)}, ~~~~\epsilon_i= \dfrac{\widehat{h}_1(X_i)\delta_i}{\widehat{S}_c(Y_i)},\\
    \widehat{w}(Y_i)& = \dfrac{1}{\sum_{j=1}^n I(Y_j>Y_i)}\sum_{j=1}^n \epsilon_j I(Y_j>Y_i), ~~~\beta_i= \widehat{w}(Y_i)(1-\delta_i),\\
     V_i &= \epsilon_i+ \beta_i- \sum_{j=1}^n \dfrac{\beta_j I(Y_i>Y_j)}{ \sum_{i=1}^n I(Y_i\geq Y_j)}~~~ \text{and}~~~\Bar{V}= \dfrac{1}{n}\sum_{i=1}^n V_i.
\end{align*}

\begin{corollary}\label{cor2}
Given that the conditions of Theorem \ref{thmcen} are satisfied, let \( \sigma^2_{0c} \) denote the value of \( \sigma^2_{c} \) under the null hypothesis \( H_0 \). As \( n \to \infty\), \( \sqrt{n}\widehat{\Delta}_{I_{c}} \) converges in distribution to a normal random variable with mean zero and variance \( 4\sigma^2_{0c} \).
\end{corollary}

Using corollary \ref{cor2}, we find the normal-based critical region of the test.  
The null hypothesis $H_0$ is rejected in favor of the alternative hypothesis $H_1$ at a significance level of $\gamma$ if

 \begin{equation*}
     \dfrac{\sqrt{n}|\widehat{\Delta}_{I_c}|}{\widehat{\sigma}_{0c}}> Z_{\gamma/2}.
 \end{equation*}
  
Section \ref{sec5} provides the results of a Monte Carlo simulation study that is used to evaluate the finite sample performance of the test.

\section{$\Delta_{M}$ : Test statistics}\label{sec4}
This section constructs a second test based on ${L}^2$ distance.

\subsection{ Uncensored case}\label{sec4.1}
The test statistic is given by  
\begin{small}
\begin{equation}\label{eqT2}
    \Delta_{M} = \int_{1}^{\infty} \Big(E\Big[\Big(\dfrac{\alpha+1}{X}\Big)\big(\min(X,t)-1 \big)\Big]- F(t)  \Big)^2 dF(t).
\end{equation}
\end{small}
Based on Theorem \ref{thm1}, $\Delta_{M}$ is non-negative under the alternative hypothesis $H_1$ and zero under the null hypothesis $(H_0)$. Now we express $\Delta_{M}$ in an alternative form. 
\begin{small}
\begin{align}\label{J(F)}
   \Delta_{M}&= \int_{1}^{\infty}E^2\Big[\Big(\dfrac{\alpha+1}{X}\Big)\big(\min(X,t)-1 \big)\Big]dF(t)\nonumber\\ &\hspace{0.3cm} -2\int_{1}^{\infty}E\Big[\Big(\dfrac{\alpha+1}{X}\Big)\big(\min(X,t)-1 \big)\Big]F(t) dF(t)
    + \int_{1}^{\infty} F^2(t) dF(t) \nonumber\\
    &= \Delta_1- \Delta_2+\Delta_3  ~~ (say).
\end{align}
\end{small}
Consider   
\begin{small}
\begin{align}\label{D1}
    \Delta_1 &= \int_{1}^{\infty}E^2\Big[\Big(\dfrac{\alpha+1}{X}\Big)\big(\min(X,t)-1 \big)\Big]dF(t)\nonumber\\
    &= (\alpha+1)^2\int_{1}^{\infty} \int_{1}^{\infty}\int_{1}^{\infty}  \Big(\dfrac{1}{xy}\Big)\big(\min(x,t)-1 \big)  \big(\min(y,t)-1 \big)dF(x)dF(y) dF(t) \nonumber\\ 
    &= (\alpha+1)^2 E \Bigg[\dfrac{\big(\min(X_1,X_3)-1 \big)  \big(\min(X_2,X_3)-1 \big)}{X_1X_2}  \Bigg],
\end{align}

\begin{align}\label{D2} 
    \Delta_2 &= 2\int_{1}^{\infty}E\Big[\Big(\dfrac{\alpha+1}{X}\Big)\big(\min(X,t)-1 \big)\Big]F(t) dF(t)\nonumber\\ 
    &= \int_{1}^{\infty}\int_{1}^{\infty} \Big(\dfrac{\alpha+1}{x}\Big)\big(\min(x,t)-1 \big) 2F(t)dF(x) dF(t)\nonumber\\ 
    &= \int_{1}^{\infty}\int_{1}^{\infty} \Big(\dfrac{\alpha+1}{x}\Big)\big((xI(x<t)+tI(t<x))-1 \big) 2F(t)dF(x) dF(t)\nonumber\\ 
     &= (\alpha+1)\int_{1}^{\infty}\int_{1}^{\infty} \bigg( I(x<t)+ \dfrac{tI(t<x)}{x} \bigg) 2F(t)dF(x) dF(t)-E\Big(\dfrac{(\alpha+1) }{X_1}\Big)\nonumber\\    
      &= (\alpha+1) \bigg(\int_{1}^{\infty}\int_{1}^{t} 2F(t)dF(x)dF(t)\nonumber\\ &\hspace{2.0cm}+\int_{1}^{\infty}\int_{1}^{\infty} \dfrac{t}{x}I(t<x) 2F(t)dF(x) dF(t) \bigg) -E\Big(\frac{(\alpha+1) }{X_1}\Big)  \nonumber\\
       &= (\alpha+1) \bigg(\frac{2}{3}+\int_{1}^{\infty}\int_{1}^{\infty}\frac{t}{x}I(t<x)  dF(x)2F(t)dF(t)  \bigg)  -E\Big(\dfrac{(\alpha+1) }{X_1}\Big)  \nonumber\\
        &= (\alpha+1) \Bigg[\frac{2}{3}+ E\Big(\dfrac{\max(X_1,X_2)}{X_3}I(\max(X_1,X_2)<X_3)\Big) \Bigg]-E\Big(\dfrac{(\alpha+1) }{X_1}\Big),  
\end{align}
\end{small}
and
\begin{align}\label{D3}
    \Delta_3 &= \int_{1}^{\infty} F^2(t) dF(t) = \dfrac{1}{3}.
\end{align}
Substituting $(\ref{D1})- (\ref{D3})$ in $(\ref{J(F)})$, we obtain 
\begin{align}\label{D2est}    
\Delta_{M}&= (\alpha+1)^2 E \Bigg[\dfrac{\big(\min(X_1,X_3)-1 \big)  \big(\min(X_2,X_3)-1 \big)}{X_1X_2}  \Bigg] \nonumber\\
&\hspace{0.5cm} -(\alpha+1) \Bigg[\frac{2}{3}+ E\bigg(\dfrac{\max(X_1,X_2)}{X_3}I\big(\max(X_1,X_2)<X_3\big)\bigg) \Bigg]\nonumber\\
&\hspace{0.5cm} + E\bigg(\dfrac{(\alpha+1)}{X_1} \bigg) +\dfrac{1}{3} \nonumber \\
&= (\alpha+1)^2 T_1- (\alpha+1)\Big(T_2 -T_3\Big) - \dfrac{(2\alpha+1)}{3} ~~~(say).
\end{align}
Hence, the test statistic is obtained using the theory of U-statistics. We consider the U-statistic defined by
\begin{equation*}
    U_{r} = \binom{n}{3}^{-1} \sum_{i=1  }^{n} \sum_{j < i, j=1}^{n} \sum_{k < j, k=1}^{n} h_r(X_i, X_j, X_k), \hspace{0.4cm} r=1,2,
\end{equation*}
and 
\begin{equation*}
    U_{3} = \binom{n}{1}^{-1} \sum_{i=1  }^{n}  h_3(X_i),
\end{equation*}
 where $h_1, h_2$ and $h_3$ are the symmetric kernels given by 
 \begin{align*}
    h_1(X_1, X_2, X_3)&= \dfrac{1}{3} \Bigg[\dfrac{ 
 \big(\min(X_1,X_3)-1 \big)  \big(\min(X_2,X_3)-1 \big)}{X_1X_2}\\
&\hspace{0.9cm} +\dfrac{\big(\min(X_1,X_2)-1 \big)  \big(\min(X_2,X_3)-1 \big)}{X_1X_3}\\
 &\hspace{0.9cm}+\dfrac{\big(\min(X_1,X_2)-1 \big)  \big(\min(X_1,X_3)-1 \big)}{X_2X_3}  \Bigg],
\end{align*}

\begin{align*}
    h_2(X_1, X_2, X_3) &=\dfrac{1}{3} \Bigg[ \dfrac{  \max (X_1,X_2)}{X_3}I(\max (X_1,X_2)\leq X_3)\\
     &\hspace{0.9cm}+\dfrac{ \max (X_1,X_3)}{X_2}I(\max (X_1,X_3)\leq X_2) \\
    &\hspace{0.9cm} +\dfrac{ \max (X_2,X_3)}{X_1}I(\max (X_2,X_3)\leq X_1) \Bigg],
\end{align*}
and 
\begin{equation*}
    h_3(X_1)=  \frac{1}{X_1}.
\end{equation*}

Note that $U_1$, $U_2$ and $U_3$ are an unbiased estimators of $T_1$, $T_2$ and $T_3$, respectively.
Hence the test statistic is given by 
\begin{equation}\label{}
    \widehat{\Delta}_{M}= (\widehat{\alpha}+1)^2U_1 -(\widehat{\alpha}+1)(U_2-U_3)-\dfrac{(2\widehat{\alpha}+1)}{3}.
\end{equation}

The test procedure is to reject the null hypothesis $H_0$ in favor of the alternative hypothesis $H_1$ for a large value of $\widehat{\Delta}_{M}$.

Next, the asymptotic properties of the test statistics are examined. According to \cite{lehmann1951consistency}, $\widehat{\alpha}$, $U_1$, $U_2$, and $U_3$ are consistent estimators of $\alpha$, $T_1$, $T_2$, and $T_3$, respectively, as they are U-statistics. Therefore,  we obtained the following result.

\begin{theorem}\label{thm5}
    Let $\widehat{\alpha}$ be the consistent estimator of $\alpha$. Under $H_1$, as $n \to \infty$, $\widehat{\Delta}_{M}$ converges in probability to $\Delta_{M}$. 
\end{theorem}

\begin{theorem}\label{thm6}
    As $\widehat{\alpha}$ be the consistent estimator of $\alpha$. The distribution of $\sqrt{n}(\widehat{\Delta}_{M}-\Delta_{M})$ converges to a normal random variable with mean zero and variance $9\sigma^2$ as $n \to \infty$, where $\sigma^2$ is obtained by
    \begin{equation*}
    \sigma^2= Var[E(h(X_1, X_2,X_3)|X_1)]. 
\end{equation*}

\begin{proof}
    Define 
     \begin{equation*}
         \widetilde{\Delta}_{M} = (\alpha+1)^2U_1 -(\alpha+1)(U_2-U_3)-\dfrac{(2 {\alpha}+1)}{3}. 
     \end{equation*}
      It is observed that $\widetilde{\Delta}_{M}$ is a U-statistic with a symmetric kernel defined as
    \begin{small}
     \begin{align}\label{h_m}
  h(X_1, X_2, X_3)
        &=\dfrac{(\alpha+1)^2}{3}  \Bigg[\dfrac{\big(\min(X_1,X_3)-1 \big)  \big(\min(X_2,X_3)-1 \big)}{X_1X_2} \nonumber\\
          &\hspace{2.3cm}+\dfrac{\big(\min(X_1,X_2)-1 \big)  \big(\min(X_2,X_3)-1 \big)}{X_1X_3}\nonumber\\
         &\hspace{2.3cm}+ \dfrac{\big(\min(X_1,X_2)-1 \big)  \big(\min(X_1,X_3)-1 \big)}{X_2X_3}  \Bigg] \nonumber\\
      &\hspace{0.5cm} -\dfrac{(\alpha+1)}{3}\Bigg[ \dfrac{  \max (X_1,X_2)}{X_3}I(\max (X_1,X_2)\leq X_3) - \frac{1}{X_1} \nonumber\\
      &\hspace{2.3cm}+\dfrac{ \max (X_1,X_3)}{X_2}I(\max (X_1,X_3)\leq X_2)-\frac{1}{X_2} \nonumber\\
    &\hspace{2.3cm} +\dfrac{ \max (X_2,X_3)}{X_1}I(\max (X_2,X_3)\leq X_1)-\frac{1}{X_3} \Bigg]\nonumber\\
    &\hspace{0.5cm}-\dfrac{(2 {\alpha}+1)}{3}.
     \end{align}
\end{small}
Consider
\begin{align*}
    \sqrt{n}(\widehat{\Delta}_{M}- \Delta_{M})&= \sqrt{n}(\widehat{\Delta}_{M}- \widetilde{\Delta}_{M})+\sqrt{n}(\widetilde{\Delta}_{M}- \Delta_{M}). 
\end{align*}
Since $\widehat{\alpha}$ be the consistent estimator of $\alpha$. We know
\begin{equation*}
    \widehat{\alpha}\xrightarrow{P} \alpha, ~ U\xrightarrow{P} E(U) \implies \widehat{\alpha} U \xrightarrow{P} \alpha E(U) \implies (\widehat{\alpha}-\alpha) U \xrightarrow{P} 0.
\end{equation*}
Using Chebyshev's inequality, $\sqrt{n}(\widehat{\Delta}_{M}- \widetilde{\Delta}_{M}) \xrightarrow{P} 0$.  
 The asymptotic distribution of $\sqrt{n}(\widetilde{\Delta}_{M}- \Delta_{M})$ is normal with mean $0$ and variance $9\sigma^2$, by central limit theorem for U-statistics (see Theorem $1$, Chapter $3$ of \cite{lee2019u}).
 So using Slutsky's theorem, we get $ \sqrt{n}(\widehat{\Delta}_{M}- \Delta_{M})\xrightarrow{d} N(0,9\sigma^2)$
\end{proof}
\end{theorem}
It should be noted that $\Delta_{M}=0$ under the null hypothesis $H_0$. Hence, the following result is obtained. 

 \begin{corollary}\label{corl2}
Under $H_0$, as $n \to \infty$, $\sqrt{n}\widehat{\Delta}_{M}$ converges in distribution to a normal random variable mean zero and variance $9\sigma_{0}^2$.
\end{corollary}

Corollary \ref{corl2} provides the asymptotic critical region for the scale-invariant test. Assuming that $\widehat{\sigma}_0^2$ is a consistent estimator of $\sigma_0^2$, the null hypothesis $H_0$ is rejected in favor of the alternative hypothesis $H_1$ at a significance level of $\gamma$ if
\begin{equation*}
    \dfrac{\sqrt{n}\widehat{\Delta}_{M}}{3\widehat{\sigma}_0}> Z_{\gamma}.
\end{equation*}
The parametric bootstrap approach is used to identify the critical point because the variance expression is not in closed form, making it difficult to determine the null variance $\sigma_0^2$.
 The critical point $(C_3)$ is determined such that $P(\widehat{\Delta}_{M}> C_3)= \gamma$. 

\subsection{ Right censored case }\label{sec4.2}
The inclusion of censored observations in the proposed testing method is now addressed. The test discussed in Section \ref{sec4.1} is then modified based on $n$ independent and identically distributed observations {$(Y_i, \delta_i), 1 \leq i \leq n $} drawn from $(Y, \delta)$.  
In the context of U-statistics theory applied to right-censored data, the same departure measure $\Delta_{M}$, as specified in (\ref{D2est}), is used.
Here, we consider the U-statistics for the right censored data given by 
\begin{align*}
     \widehat{\Delta}^*_{M_{cr}}  &= \dfrac{2}{n(n-1)(n-2)}\sum_{i=1  }^{n} \sum_{j < i, j=1}^{n}\sum_{k < j, k=1}^{n} \dfrac{h^*_{1r}(Y_i, Y_j, Y_k)\delta_i \delta_j\delta_k }{\widehat{S}_c(Y_i)\widehat{S}_c(Y_j)\widehat{S}_c(Y_k)}, ~~ r= 1, 2, 
\end{align*}
and
\begin{align*}
     \widehat{\Delta}^*_{M_{c3}} &= \dfrac{1}{n}\sum_{i=1  }^{n} \Big(\frac{1}{Y_i}\Big)\dfrac{\delta_i}{\widehat{S}_c(Y_i)}. 
\end{align*}

The symmetric kernels for $\widehat{\Delta}^*_{M_{c1}}$ and $\widehat{\Delta}^*_{M_{c2}}$ are then

\begin{align*}
    h^*_{11}(Y_i, Y_j, Y_k)&= \Bigg[\dfrac{ 
 \big(\min(Y_i,Y_k)-1 \big)  \big(\min(Y_j,Y_k)-1 \big)}{Y_iY_j}\\
 &\hspace{0.6cm}+\dfrac{ 
 \big(\min(Y_i,Y_j)-1 \big)  \big(\min(Y_j,Y_k)-1 \big)}{Y_iY_k}\\
 &\hspace{0.6cm}+\dfrac{\big(\min(Y_i,Y_j)-1 \big)  \big(\min(Y_i,Y_k)-1 \big)}{Y_jY_k} \Bigg],
\end{align*}
and
\begin{align*}
    h^*_{12}(Y_i, Y_j, Y_k) &=\Bigg[ \dfrac{  \max (Y_i,Y_j)}{Y_k}I(\max (Y_i,Y_j)\leq Y_k) \\
    &\hspace{0.5cm}+\dfrac{ \max (Y_i,Y_k)}{Y_j}I(\max (Y_i,Y_k)\leq Y_j) \\
    &\hspace{0.5cm} +\dfrac{ \max (Y_j,Y_k)}{Y_i}I(\max (Y_j,Y_k)\leq Y_i)  \Bigg],
\end{align*}
respectively.  
The moment-based estimator of $\alpha(>1)$ under the censored case is obtained as
\begin{equation*}
   \Bar{X}_c = \dfrac{1}{n} \sum_{i=1  }^{n} \dfrac{Y_i \delta_i} {\widehat{S}_c(Y_i)} ~~ \text{and} ~~ \widehat{\alpha}_c = \dfrac{\Bar{X}_c}{(\Bar{X}_c-1)}, 
\end{equation*}
where $\Bar{X}_c$ denotes the censored sample mean of the random variable $X$. 
Since $\widehat{\alpha}_c$ is consistent estimator of $\alpha$.
Hence, the test statistic is obtained as
\begin{equation}
    \widehat{\Delta}_{M_c} = ( \widehat{\alpha}_c+1)^2 \widehat{\Delta}^*_{M_{c1}}  -(\widehat{\alpha}_c+1) ( \widehat{\Delta}^*_{M_{c2}} - \widehat{\Delta}^*_{M_{c3}} )-\dfrac{(2\widehat{\alpha}_c+1)}{3}. 
\end{equation}

Define $N_i^c(t) = I(Y_i \leq t, \delta_i = 0)$ as the counting process corresponding to the censoring random variable for the $i$-th subject and $R_i(t) = I(Y_i \geq t)$ to obtain the asymptotic distribution of $\widehat{\Delta}_{M_c}$. Furthermore, let $\alpha_c(\cdot)$ be the hazard function corresponding to the censoring variable $C$. Given this counting process $N_i^c(t)$, the martingale associated with it is given by
\begin{equation*}
    M_i^c(t)= N_i^c(t)- \int_{0}^{t} R_i(u)\alpha_c(u) du, ~~ i=1,\ldots,n. 
\end{equation*}
Now, consider the kernel function conditioning on $Y_1=x$ as
\begin{align*}
    h_2(x) &=\dfrac{(\alpha+1)^2}{3} E \Bigg[\dfrac{\big(\min(x,Y_3)-1 \big)  \big(\min(Y_2,Y_3)-1 \big)}{xY_2}\nonumber\\
         &\hspace{2cm} +\dfrac{\big(\min(x, Y_2)-1 \big)  \big(\min(Y_2,Y_3)-1 \big)}{xY_3}\nonumber\\
         &\hspace{2cm}+ \dfrac{\big(\min(x, Y_2)-1 \big)  \big(\min(x, Y_3)-1 \big)}{Y_2Y_3}  \Bigg] \nonumber\\
      &\hspace{0.5cm} -\dfrac{(\alpha+1)}{3} E\Bigg[ \dfrac{  \max (x, Y_2)}{Y_3}I(\max (x, Y_2)\leq Y_3) - \frac{1}{x} \\
       &\hspace{2.3cm}+\dfrac{ \max (x, Y_3)}{Y_2}I(\max (x, Y_3)\leq X_2)-\frac{1}{Y_2} \\
    &\hspace{2.3cm} +\dfrac{ \max (Y_2, Y_3)}{x}I(\max (Y_2, Y_3)\leq x)-\frac{1}{Y_3} \Bigg] +\frac{1}{3},
     \end{align*}
such that $E(h(Y_1, Y_2, Y_3)|Y_1=x)= h_2(x)$, for proof see \cite{datta2010inverse}. Let $H_c(x,t)= P(X_1\leq x, Y_1\leq t, \delta=1),~ x \in \mathcal{X}$,~ $t\geq 0$, be the empirical sub-distribution function of the pair $(X_i, Y_i)$, $\Bar{K}(t)=P(Y_1>t)$ be the survival function and the weight function is given by
\begin{equation*}
    w(t)= \dfrac{1}{\Bar{K}(t)} \int_{\mathcal{X}\times [0, \infty)} \dfrac{h_2(x)}{S_c(y)} I(y>t)dH_c(x,y), ~~ t\geq 0.
\end{equation*}

Next, we examined the asymptotic properties $\sqrt{n}(\widehat{\Delta}_{M_c}) $ and the proof of the following theorem follows a similar approach to that of Theorem $6$, using a specific choice of the kernel function (see \cite{datta2010inverse}).
\begin{theorem}\label{thmcen2}
Assume that
\begin{small}
\begin{equation*}
    E\big[ h(Y_1, Y_2, Y_3)  \big]^2 <\infty, ~ \int_{\mathcal{X}\times [0, \infty)}\dfrac{h^2_2(x)}{S^2_c(y)}dH_c(x,y)<\infty ~~ \text{and} ~~\int_0^\infty w^2(t) \alpha_c(t) dt<\infty.  
\end{equation*}
\end{small}
As $n \to \infty$, $\sqrt{n}(\widehat{\Delta}_{M_c} - \Delta_{M}) $ converges to a normal random variable with mean zero and variance $9\sigma^2_{c}$, where $\sigma^2_{c}$ is given by
\begin{equation*}
    \sigma^2_{c}= Var\Big( \dfrac{h_2(X)\delta_1}{S_c(Y_1)}+\int w(t) dM_1^c(t) \Big). 
\end{equation*}
\end{theorem}
Reweighed approaches are employed to obtain an estimator for \( \sigma^2_{c} \).
Consider

\begin{align*}
    \widehat{h}_2(x)&=  \dfrac{1}{n^2}\sum_{j,k=1}^n  \dfrac{h(x,Y_j, Y_k)\delta_j\delta_k}{\widehat{S}_c(Y_j)\widehat{S}_c(Y_k)}, ~~~~\epsilon_i= \dfrac{\widehat{h}_2(X_i)\delta_i}{\widehat{S}_c(Y_i)},\\
    \widehat{w}(Y_i) &= \dfrac{1}{\sum_{j,k=1}^n  I(Y_j>Y_i)I(Y_k>Y_i)}\sum_{j,k=1}^n\epsilon_j \epsilon_k I(Y_j>Y_i)I(Y_k>Y_i),\\
    \Bar{V}&= \dfrac{1}{n}\sum_{i=1}^n V_i, ~~~
    \beta_i = \widehat{w}(Y_i)(1-\delta_i) \\ 
    \text{and}~~ V_i &= \epsilon_i+ \beta_i - \sum_{j=1}^n \dfrac{\beta_j I(Y_i>Y_j)}{ \sum_{i=1}^n I(Y_i\geq Y_j)} -\sum_{k=1}^n \dfrac{\beta_k I(Y_i>Y_k) }{ \sum_{i=1}^n I(Y_i\geq Y_k)}.
\end{align*}

 Hence
\begin{equation*}
    \widehat{\sigma}^2_{c}= \dfrac{9}{(n-1)} \sum_{i=1}^n (V_i- \Bar{V})^2. 
\end{equation*}

\begin{corollary}
Under the assumptions specified in Theorem \ref{thmcen2}, let \( \sigma^2_{0c} \) denote the value of \( \sigma^2_{c} \) when evaluated under the null hypothesis \( H_0 \). As \( n \to \infty\), \( \sqrt{n}\widehat{\Delta}_{M_c} \) will converge in distribution to a normal random variable with mean zero and variance \( 9\sigma^2_{0c} \) under \( H_0 \).
\end{corollary}

At a significance level $\gamma$, the null hypothesis $H_0$ is rejected in favor of the alternative hypothesis $H_1$ in the setting of right-censored data if
 \begin{equation*}
     \dfrac{\sqrt{n}\widehat{\Delta}_{M_c}}{3\widehat{\sigma}_{0c}}> Z_{\gamma}.
 \end{equation*}
 
 The performance of the test with a finite sample is evaluated using a Monte Carlo simulation study, and the results are presented in Section \ref{sec5}.
 
\section{Simulation study and results}\label{sec5}
In this section, Monte Carlo simulations are employed to assess and compare the finite-sample performance of the newly suggested tests with the current Pareto type-I distribution tests.

\begin{itemize} 
        \item Given a non-negative random variable $X$ with a common distribution function $F$, let $X_1, \ldots, X_n$ be identical copies of it. For any integer $m$, let $2\leq m\leq n$. Then the distribution of the random variables $X^{m^{-1}}$ and $\min \{X_1, \ldots, X_m\}$ is the same if, and only if, for each $t\in \mathcal{R}$,
         
        \begin{equation*}
            E\Big\{ \dfrac{1}{m} \exp \Big(-it X^{m^{-1}}\Big) -[1-F(X)]^{m-1} \exp(-itX)  \Big\}= 0,
        \end{equation*}
where $F(X)$ is the distribution function of the Pareto type-I distribution, which is given in (\ref{eq1}). Recently, \cite{ngatchou2024classes} proposed the following test statistic
\begin{equation*}
    T_{m,n,w}= \int_{\mathcal{R}} |S_{m,n,\widehat{\alpha}_n}(t) |^2 w(t) dt,
\end{equation*}
where for all $t\in\mathcal{R}$,
\begin{equation*}
    S_{m,n,\alpha}(t)= \dfrac{1}{\sqrt{n}}\sum_{j=1}^n \Big[ \dfrac{1}{m}\exp(-itX_j^{m^{-1}})-X_j^{-\alpha(m-1)}\exp(-itX_j)\Big].
\end{equation*}
      Let us consider the parameter used in the weight function as $a=0.5$ for the practical application. The two test statistics introduced 
\begin{enumerate}
     \item $T_{m,a}^{(1)}$ - based on Laplace weight function,  $w(t)= e^{-a|t|}$.
     \item $T_{m,a}^{(2)}$ - based on normal weight function, $w(t)= e^{-at^2}$.
\end{enumerate}

        \item Based on likelihood ratio, \cite{zhang2002powerful} proposed two tests with test statistics given by
        \begin{equation*}
            ZA_n= -\sum_{j=1}^n \Bigg[\dfrac{\log \big(1-F(X_{(j)})^{-1}\big)}{n-j+0.5}+ \dfrac{\log \big(F(X_{(j)})^{-1}\big)}{j-0.5} \Bigg]
        \end{equation*}
and 
\begin{equation*}
            ZB_n= \sum_{j=1}^n \Bigg[ \log \Bigg( \dfrac{ \big(1-F(X_{(j)})^{-1} \big)^{-1} -1 }{  (n-0.5)/(j-0.75)-1}   \Bigg) \Bigg]^2,
        \end{equation*}
where the $i$-th order statistic based on a random sample $X_1, \ldots, X_n$ from $F$ is denoted by $X_{(i)}$.
        
          \item \cite{meintanis2009goodness} proposed a test based on the empirical distribution function. The test statistics based on the transformation $\widehat{U}_j= F(X_j);$ $ j=1, \ldots, n$, is 
          \begin{align*}
              ME_n&= \dfrac{1}{n}\sum_{j,k=1}^n \dfrac{2a}{(\widehat{U}_j-\widehat{U}_k)^2 +a^2} -4\sum_{j=1}^n \Big[\tan^{-1}\Big(\dfrac{\widehat{U}_j}{a}\Big)+\tan^{-1}\Big(\dfrac{1-\widehat{U}_j}{a} \Big)\Big] \\
              &\hspace{0.3cm} +2n \Big[2\tan^{-1} \Big(\dfrac{1}{a}\Big)-a\log \Big(1+\dfrac{1}{a^2} \Big) \Big].
          \end{align*}
The tuning parameter is set to $a = 0.5$ to generate the presented Monte Carlo results.

            \item  Let $X$ and $Y$ be independent and identical positive continuous random variables. The distribution of the random variables $X$ and $\max \Big\{\dfrac{X}{Y},\dfrac{Y}{X}\Big\}$ are identical if, and only if, $X$ has a Pareto type-I distribution. Based on this characterization, the tests are provided in \cite{obradovic2015goodness}. The test statistic is
        \begin{align*}
            OJ_n &= \int_0^\infty (M_n(x)-F_n(x)) dF_n(x),
        \end{align*}
                
            where $M_n(x)= \binom{n}{2}^{-1} \sum_{i=1  }^{n} \sum_{j < i, j=1}^{n} I \Big\{\max \Big(\dfrac{X}{Y},\dfrac{Y}{X} \Big) \leq x \Big\}, ~~ x\geq 1$,
            and $F_n(x)= \dfrac{1}{n} \sum_{i=1}^n I\{X_i\leq x\}$.

             \item Suppose there exists a distribution function $F$ such that $X, X_1,\ldots, X_n$ are independent, identical and positive continuous random variables.  For all integers $2\leq m\leq n$, random variables \(\sqrt[m]{X}\) and \(\min(X_1,\ldots, X_m)\) have the same distribution if, and only if, $F$ is the Pareto type-I distribution. Following from this characterization, \cite{allison2022distribution} suggested three tests for the Pareto distribution. The test statistics are given by 
             \begin{align*}
                  I_{n,m}&= \int_1^\infty \Delta_{n,m}(x)dF_n(x),\\
                 \text{and} ~~~ M_{n,m}&= \int_1^\infty \Delta^2_{n,m}(x)dF_n(x), 
             \end{align*}
             whereas the discrepancy measures of $\sqrt[m]{X}$, defined as 
             \begin{small}
             \begin{equation*}
                 \Delta_{n,m}(x)= \dfrac{1}{n}\sum_{j=1}^n I \Big\{X_{j}^{m^{-1}} \leq x\Big\}- \dfrac{1}{n^m} \sum_{j_1,\ldots, j_m=1}^n I\{\min(X_{j_1}, \ldots, X_{j_m}) \leq x \}. 
             \end{equation*}
              \end{small}
             To generate the presented Monte Carlo results, the tuning parameter is set to \(m = 2\).

    \item Cramér–von Mises (CvM) test statistic,
 
 \hspace{2.7cm}    $CvM$= $\int \big(F_n(x)-{F}(x) \big)^2 dF(x)$.

     The above test statistic can be expressed using the order statistics as
    \begin{equation*}
        CvM= \sum_{i=1}^n \Big[ F(X_{(i)})- \dfrac{2i-1}{2n} \Big]^2 +\dfrac{1}{12n}, 
    \end{equation*}
where $X_{(j)}$ denotes the order statistics.  
   
    \item Anderson-Darling (AD) test statistic
    \begin{equation*}
        AD= \int \dfrac{\big(F_n(x) - {F}(x)\big)^2}{F(x)(1-F(x))}dF(x). 
    \end{equation*}
The above test statistic can be formulated using the order statistics as 
     \begin{equation*}
         AD= -n-\dfrac{1}{n}\sum_{i=1}^n (2i-1) \big[\log \big(F(X_{(i)})\big)+ \log (1- F(X_{(n+1-i)})) \big].
     \end{equation*}

\item  Kolmogorov-Smirnov (KS) test statistic 
\begin{equation*}
    KS= \sup_{x\geq1} \big|F_n(x)-F(x)\big|.
\end{equation*}
\end{itemize}

The Monte Carlo approach with $10,000$ replications at the $0.05$ significance level is used to estimate the empirical critical values for all tests. The parameter of the Pareto type-I distribution is obtained using the moment-based estimator, $\widehat{\alpha}= \Bar{X}/(\Bar{X}-1)$. To estimate the empirical sizes and powers of the proposed tests, sample sizes of $n = 25, 50, 75$, and $100$ are used. All computations and simulations are exclusively carried out using R software. A wide range of alternative distributions is considered and presented in Table \ref{table:1} for comparison purposes. 
\begin{small}
\begin{figure}[h!]
    \centering
    \includegraphics[width=13cm, height=8cm]{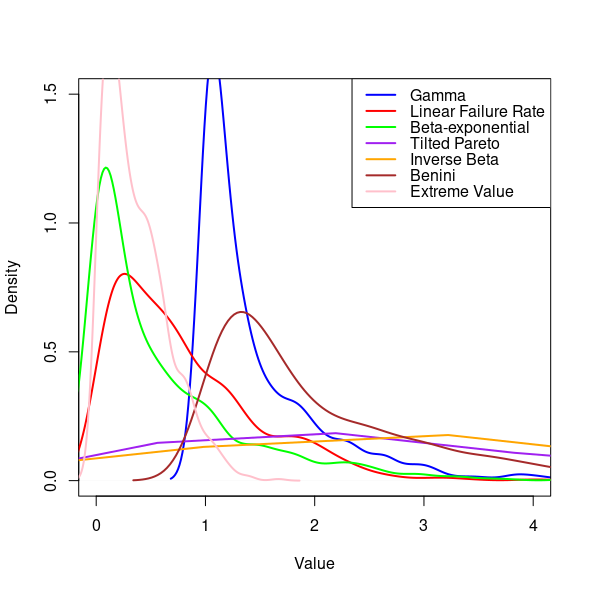}
    \caption{Density plots of alternative distributions $(n=1000, \lambda = 0.5)$}
    \label{fig:enter-label}
\end{figure}
\end{small}

\begin{small}
\begin{table}[h!]
\centering
\caption{Lists of alternative distributions}
\vspace{0.2cm}
\label{table:1} 
\fontsize{9pt}{11pt}\selectfont
\begin{tabular}{p{4cm} p{7cm} p{2cm} } 
\hline 
 Distribution & Form of density function & Notation  \\[1 ex]
\hline 
Gamma &$ ({\Gamma(\lambda)})^{-1}(x-1)^{\lambda-1}e^{-(x-1)} $  & $\Gamma(\lambda)$ \\[0.5ex]
Linear failure rate & $(1+\lambda(x-1))\exp(-(x-1)-\lambda(x-1)^2/2)$ & $LF(\lambda)$\\[0.5ex]
Beta exponential & $\lambda e^{-(x-1)}(1-e^{-(x-1)})^{\lambda-1}$ & $BE(\lambda)$\\[0.5ex]
Tilted Pareto&$ (1+\lambda)(x+\lambda)^{-2}$  & $TP(\lambda)$\\[0.5ex]
Inverse beta& $(1+\lambda)(x-1)^\lambda x^{-(2+\lambda)}$& $IB(\lambda)$\\[0.5ex]
Benini  & $x^{-2} (1+2\lambda \ln x) e^{-\lambda \ln^2 x}$ & $B(\lambda)$\\[0.5ex]
Extreme value&  $\lambda^{-1} \exp(-x/\lambda) \exp(-\exp(-x/\lambda)) $  & $EV(\lambda)$ \\[0.5ex]
\hline
\end{tabular}
\end{table}
\end{small}

The parametric bootstrap method is a powerful statistical tool for estimating critical points in various hypothesis testing scenarios. By generating multiple resamples from a fitted parametric model, this approach allows for a robust assessment of the distribution of the test statistic under the null hypothesis. The critical point, which is pivotal in deciding whether to reject the null hypothesis, is identified through this resampling procedure. The specific algorithm utilized in this study is outlined in Algorithm $1$, providing a clear and systematic approach to applying the parametric bootstrap in practice. 

\begin{algorithm} 
\caption{An algorithm to find the $C_1$ and $C_2$}\label{alg:cap}
\begin{algorithmic}
\State  $x$ is a numeric vector of data values
\State  $\Bar{X} \gets$ mean$(x)$ 
\State $n \gets$ length$(x)$
\State $\widehat{\alpha}\gets \Bar{X}/(\Bar{X} -1) $  \Comment{Estimation of parameter}
\State delta($x, \widehat{\alpha}$)   \Comment{Calculate the test statistic}
\State $B\gets 1000$   \Comment{No. of bootstrap replicates}
\State for $(b ~\text{in}~ 1:B )$\{
\State $ i \gets$ sample(1:n, size=n, replicate=TRUE)   
\State  $ y \gets x[i]$
\State deltas[b] $\gets$ delta $(y, \widehat{\alpha})$ \}
\State deltas $\gets$ sort(deltas)
\State $C_1 \gets$ quantile(deltas, $\gamma/2$) \Comment{lower bound}
\State $C_2 \gets$ quantile(deltas, $1-\gamma/2$)  \Comment{upper bound}
\State ifelse((delta $< C_1||$delta $> C_2$), print("Reject $H_0$"), print("Accept $H_0$"))
\end{algorithmic}
\end{algorithm}
The algorithm employs the parametric bootstrap method to estimate the test statistic by generating $1,000$ resampled datasets, computing the test statistic for each sample, and deriving critical bounds $(C_1, C_2)$ from the empirical distribution of these statistics. The null hypothesis \(H_0\) is rejected if the observed test statistic lies beyond these critical thresholds.

The results of the simulation study are reported in Tables \ref{table:2}–\ref{table:5}. It is observed from these tables that for both tests, the empirical type I error converges to the given level of significance. The analysis reveals that, in the majority of cases evaluated, the newly proposed tests exhibit superior performance compared to other tests.
For the censored case, for finding the power, the lifetime random variable is generated from the same alternative reported in Table \ref{table:1}. In all instances, the censoring random variable is generated from an exponential distribution with a rate parameter $b$. For a sample containing $20\%$ and $40\%$ censored observations, the rate parameter $b$ is calculated such that $P(X > C) = 0.2$ and $0.4$, respectively. The results of this simulation study are given in Tables \ref{table:6} and \ref{table:7}. It can be noticed that, even with small sample sizes, the performance of both tests remains very robust at the given level of significance.

\begin{landscape} 
\begin{table}   
\centering
\caption{Empirical size and power at 0.05 level of significance}
\label{table:2} 
\fontsize{9pt}{11pt}\selectfont
\begin{tabular}{p{1cm} p{0.6cm} p{1cm}p{1cm} p{1cm}p{1cm} p{1cm}  p{1cm}p{1cm} p{1cm} p{1cm} p{1cm} p{1cm} p{1cm} p{1cm}} 
\hline \\
      &$n$  & $\widehat{\Delta}_{I}$& $\widehat{\Delta}_{M}$&  $T^{(1)}_{3,0.5}$ & $T^{(2)}_{3,0.5}$ & $ZA_n$  & $ZB_n$  & $ME_n$   & $OJ_n$  & $I_{n,m}$   & $M_{n,m}$   & CvM &  AD & KS    \\[1ex]
\hline \\
 $P(1)$& 25  & 0.068 & 0.102&  0.057&  0.071& 0.116& 0.116& 0.101 & 0.050& 0.045&0.052& 0.106& 0.360& 0.093 \\[1ex]
        & 50  & 0.066 &0.083 & 0.061 & 0.085& 0.141& 0.143& 0.116&0.053& 0.051&0.049& 0.126& 0.346 &0.107  \\[1ex]
        & 75  & 0.053 &0.068 &0.071 & 0.086& 0.137 &0.155& 0.110&0.043& 0.039 &0.037&0.112& 0.372& 0.087  \\[1ex]
        & 100 & 0.061 &0.051 & 0.058 & 0.070 &0.186& 0.212 &0.111 &0.055&0.051 &0.026&0.112& 0.390 &0.115  \\[1.5ex]
   \hline \\    
 $P(5)$& 25  & 0.098 &0.068 & 0.046 & 0.045& 0.047& 0.043& 0.045&0.045 &0.043&0.047& 0.047& 0.317& 0.048   \\[1ex]
        & 50  & 0.073 & 0.042&  0.051  &0.053 &0.042& 0.044 &0.044 &0.056&0.047&0.051 &0.051 &0.313 &0.043   \\[1ex]
        & 75  & 0.047 &0.065  &  0.050 & 0.048& 0.072& 0.064& 0.059&0.055& 0.047&0.056& 0.051& 0.397 &0.057   \\[1ex]
        & 100 & 0.056 &0.056 &  0.030&  0.033& 0.031& 0.035& 0.031&0.029& 0.028&0.028& 0.030& 0.307 &0.037   \\[1.5ex]
   \hline \\ 
$\Gamma(0.5)$& 25  & 0.991 & 1.000 & 0.149 & 0.129& 0.113& 0.121& 0.099&0.170& 0.184&0.128& 0.106& 0.537& 0.105 \\[1ex]
             & 50  & 1.000 &1.000 &0.322 & 0.267 &0.393& 0.477& 0.290&0.481& 0.513&0.364& 0.241 &0.824 &0.233  \\[1ex]
             & 75  & 1.000 &1.000 & 0.514 & 0.423& 0.681& 0.768& 0.449&0.682& 0.746&0.638& 0.404 &0.950 &0.435   \\[1ex]
             & 100 & 1.000 &1.000 & 0.623&  0.530& 0.792& 0.848& 0.590&0.781& 0.826&0.742& 0.551& 0.971 &0.540  \\[1.5ex]
 \hline \\ 
 $\Gamma(1)$& 25  & 1.000 &1.000 & 0.457&  0.454 &0.415 &0.469 &0.487 &0.373&0.303&0.278& 0.501& 0.848 &0.426   \\[1ex]
            & 50  & 1.000 &1.000 &  0.688&  0.709& 0.735 &0.794& 0.738 &0.650&0.559&0.415& 0.760 &0.971 &0.680 \\[1ex]
            & 75  & 1.000  &1.000 &  0.884&  0.904& 0.930& 0.950& 0.939&0.822& 0.732&0.625& 0.953 &0.996 &0.883  \\[1ex]
            & 100 & 1.000  &1.000 &  0.979 & 0.980& 0.988& 0.994& 0.984&0.940& 0.875&0.773& 0.990 &1.000 &0.968   \\[1.5ex]
 \hline \\ 
 \end{tabular}
\end{table}
\end{landscape}

\begin{landscape}
\begin{table}  
\centering
\caption{Empirical power at 0.05 level of significance}
\vspace{0.2cm}
\label{table:3} 
\fontsize{9pt}{11pt}\selectfont
\begin{tabular}{p{1cm} p{0.6cm} p{1cm}p{1cm} p{1cm}p{1cm} p{1cm}  p{1cm}p{1cm} p{1cm} p{1cm} p{1cm} p{1cm} p{1cm} p{1cm}} 
\hline \\
   &$n$  & $\widehat{\Delta}_{I}$& $\widehat{\Delta}_{M}$ &  $T^{(1)}_{3,0.5}$ & $T^{(2)}_{3,0.5}$ & $ZA_n$  & $ZB_n$  & $ME_n$   & $OJ_n$  & $I_{n,m}$   & $M_{n,m}$   & CvM &  AD & KS    \\[1ex]
\hline \\
$LF(0.5)$ & 25  & 0.963 &0.983 & 0.588&  0.565& 0.617& 0.688& 0.611&0.569& 0.480&0.393& 0.634& 0.925& 0.577 \\[1ex]
          & 50  & 0.997& 1.000 &0.878&  0.876 &0.928& 0.942& 0.900 &0.857&0.773 &0.693&0.916& 0.994 &0.830       \\[1ex]
          & 75  & 1.000  &1.000 & 0.981 & 0.979& 0.993& 0.993& 0.991&0.970& 0.944&0.881& 0.994 &1.000& 0.978  \\[1ex]
          & 100 & 1.000 &1.000 & 0.999&  0.997& 1.000& 1.000 &0.999&0.995& 0.981&0.972& 1.000 &1.000 &0.999  \\[1.5ex]
 \hline \\ 
 $LF(1)$ & 25  & 0.973&0.998  & 0.700  &0.699 &0.700 &0.772 &0.728 &0.683&0.568& 0.502&0.738 &0.954 &0.653  \\[1ex]
         & 50  & 1.000  & 1.000& 0.948 & 0.932& 0.970 &0.976 &0.961&0.935& 0.862 &0.799&0.960& 0.999 &0.896  \\[1ex]
         & 75  & 1.000 &1.000 &0.991 & 0.989 &0.996 &0.998 &0.994&0.988& 0.973&0.946& 0.996 &1.000& 0.985   \\[1ex]
         & 100 & 1.000 &1.000 & 1.000  &1.000& 1.000 &1.000 &1.000&0.998& 0.989 &0.981&1.000& 1.000 &0.999   \\[1.5ex]
 \hline \\ 
 $BE(0.5)$ & 25  & 0.930 &1.000 & 0.130 & 0.095 &0.115& 0.142 &0.111&0.213& 0.186&0.147& 0.095& 0.589& 0.098\\[1ex]
           & 50  & 0.990  &0.997 &0.280&  0.175 &0.395 &0.470 &0.271 &0.382&0.404 &0.409&0.224 &0.823 &0.229      \\[1ex]
           & 75  & 1.000  & 1.000& 0.381 & 0.269& 0.618& 0.695& 0.371&0.503 &0.555&0.532& 0.286& 0.925 &0.280   \\[1ex]
           & 100 &  1.000 &1.000 & 0.494&  0.309 &0.814& 0.873&0.472&0.627& 0.686&0.643& 0.378& 0.966 &0.412  \\[1.5ex]
\hline\\
$BE(1)$ & 25  & 0.946 &0.994 & 0.425 & 0.466& 0.469 &0.506 &0.500&0.380& 0.255 &0.226&0.517 &0.881& 0.421   \\[1ex]
        & 50  & 0.992 &1.000 &  0.689 & 0.706& 0.745& 0.799& 0.801&0.618& 0.499&0.423& 0.794 &0.974 &0.713   \\[1ex]
        & 75  & 1.000  &1.000 & 0.929&  0.935 &0.944& 0.954& 0.961&0.839& 0.749&0.608& 0.961& 0.998 &0.904   \\[1ex]
        & 100 & 1.000 &1.000 & 0.978 & 0.979& 0.990 &0.993& 0.990&0.933& 0.884&0.766& 0.992& 0.999 &0.971 \\[1.5ex]
\hline
\end{tabular}
\end{table}
\end{landscape}

\begin{landscape}
\begin{table}  
\centering
\caption{Empirical power at 0.05 level of significance}
\vspace{0.2cm}
\label{table:4} 
\fontsize{9pt}{11pt}\selectfont
\begin{tabular}{p{1cm} p{0.6cm} p{1cm}p{1cm} p{1cm}p{1cm} p{1cm}  p{1cm}p{1cm} p{1cm} p{1cm} p{1cm} p{1cm} p{1cm} p{1cm}} 
\hline \\
      &$n$  & $\widehat{\Delta}_{I}$& $\widehat{\Delta}_{M}$&  $T^{(1)}_{3,0.5}$ & $T^{(2)}_{3,0.5}$ & $ZA_n$  & $ZB_n$  & $ME_n$   & $OJ_n$  & $I_{n,m}$   & $M_{n,m}$   & CvM &  AD & KS    \\[1ex]
\hline \\
 $TP(0.5)$& 25  & 0.860 & 0.946  & 0.131&  0.172& 0.378& 0.408& 0.289&0.092 & 0.074&0.083& 0.313 &0.639 &0.265\\[1ex]
        & 50  & 0.984 &1.000 &0.250 & 0.304& 0.531& 0.557& 0.470&0.128& 0.106& 0.106&0.466 &0.779 &0.416   \\[1ex]
        & 75  & 0.991 &1.000 & 0.376 & 0.449& 0.711& 0.709& 0.610&0.157& 0.133& 0.130&0.608& 0.872 &0.606    \\[1ex]
        & 100 & 1.000 &1.000 & 0.464 & 0.541 &0.786& 0.787& 0.712&0.193& 0.161 &0.118&0.723& 0.921 &0.634  \\[1.5ex]
   \hline \\     
 $TP(1)$& 25  & 0.968 &1.000 & 0.266 & 0.349 &0.631 &0.634& 0.544&0.145& 0.117&0.141& 0.553& 0.838& 0.481   \\[1ex]
        & 50  & 1.000 & 1.000& 0.494 & 0.596 &0.873& 0.885& 0.782&0.238& 0.198&0.146& 0.801& 0.957 &0.725  \\[1ex]
        & 75  & 1.000 &1.000 &   0.667 & 0.779& 0.963 &0.964& 0.904&0.349& 0.311 &0.262&0.918 &0.987 &0.866  \\[1ex]
        & 100 & 1.000 &1.000 &   0.816 & 0.886& 0.983& 0.984 &0.966 &0.428&0.396 &0.280&0.969 &0.996 &0.938  \\[1.5ex]
 \hline \\ 
$IB(0.5)$ & 25  & 0.906& 0.969& 0.250&  0.296 &0.518 &0.532 &0.406&0.165 &  0.148&0.114& 0.429& 0.724 &0.345  \\[1ex]
        & 50  & 0.995 &1.000 &0.453 & 0.564& 0.770 &0.743& 0.664& 0.280&0.240&0.213& 0.687 &0.908 &0.593   \\[1ex]
        & 75  & 1.000  &1.000 & 0.648&  0.719& 0.876 &0.867 &0.772&0.410& 0.379& 0.274&0.800 &0.940 &0.709  \\[1ex]
        & 100 & 1.000 &1.000 & 0.800  &0.866 &0.944 &0.934 &0.879&0.574& 0.527&0.355&0.885& 0.979 &0.829  \\[1.5ex]
 \hline \\ 
 $IB(1)$ & 25  & 0.891 & 0.942&  0.409 & 0.490 &0.684& 0.682& 0.629& 0.243& 0.190&0.180& 0.660& 0.877 &0.571  \\[1ex]
        & 50  & 0.940 &0.998 &  0.745 & 0.806& 0.935& 0.930& 0.904&0.525& 0.475&0.362& 0.912& 0.983& 0.883 \\[1ex]
        & 75  & 1.000 &1.000 &  0.947 & 0.971& 0.994& 0.993& 0.984 &0.786&0.768&0.610 &0.987 &0.999& 0.967  \\[1ex]
        & 100 & 1.000 &1.000 & 0.992 & 0.996& 0.999& 0.999& 0.997&0.920& 0.907& 0.782&0.997& 1.000 &0.992  \\[1.5ex]
\hline
\end{tabular}
\end{table}
\end{landscape}

\begin{landscape}
\begin{table}  
\centering
\caption{Empirical power at 0.05 level of significance}
\vspace{0.2cm}
\label{table:5} 
\fontsize{9pt}{11pt}\selectfont
\begin{tabular}{p{1cm} p{0.6cm} p{1cm}p{1cm} p{1cm}p{1cm} p{1cm}  p{1cm}p{1cm} p{1cm} p{1cm} p{1cm} p{1cm} p{1cm} p{1cm}} 
\hline \\
      &$n$  & $\widehat{\Delta}_{I}$& $\widehat{\Delta}_{M}$&  $T^{(1)}_{3,0.5}$ & $T^{(2)}_{3,0.5}$ & $ZA_n$  & $ZB_n$  & $ME_n$   & $OJ_n$  & $I_{n,m}$   & $M_{n,m}$   & CvM &  AD & KS    \\[1ex]
\hline \\
 $B(0.5)$ & 25  & 0.954 & 0.997 & 0.233&  0.237& 0.252& 0.287& 0.268&0.229&   0.157&0.150& 0.273& 0.683 &0.245  \\[1ex]
        & 50  & 0.993 &1.000 &0.513 & 0.521 &0.495 &0.514& 0.568 &0.410&0.328&0.256& 0.591& 0.865& 0.510   \\[1ex]
        & 75  & 1.000  &1.000 & 0.666&  0.707& 0.696& 0.731& 0.750&0.521& 0.479& 0.378&0.764& 0.963 &0.717  \\[1ex]
        & 100 & 1.000 &1.000 & 0.773&  0.808 &0.789 &0.824& 0.841&0.684& 0.607&0.451& 0.848 &0.985 &0.773   \\[1.5ex]
 \hline \\ 
$B(1)$ & 25  & 0.961 &1.000 & 0.414 & 0.432& 0.448& 0.464& 0.459&0.375& 0.301&0.280& 0.457& 0.827& 0.433  \\[1ex]
        & 50  & 1.000  &1.000 & 0.687 & 0.715& 0.676 &0.722 &0.712&0.628& 0.552&0.399 &0.722& 0.933& 0.663  \\[1ex]
        & 75  & 1.000 &1.000 & 0.878 & 0.870 &0.850& 0.890& 0.889&0.801& 0.731&0.542 &0.896 &0.988 &0.826 \\[1ex] 
        & 100 & 1.000 &1.000 & 0.949 & 0.940& 0.916 &0.935 &0.949&0.887& 0.840 &0.701&0.956& 0.998 &0.915   \\[1.5ex]
\hline\\
$EV(0.5)$& 25  & 1.000 &1.000 & 0.989 & 0.999& 1.000& 1.000& 1.000   &1.000 &0.998&0.994& 1.000& 1.000 &1.000  \\[1ex]
        & 50  & 1.000 &1.000 &0.997 & 0.999 &1.000& 1.000& 1.000&1.000& 1.000&1.000& 1.000& 1.000& 1.000    \\[1ex]
        & 75  & 1.000 &1.000 & 0.998&  1.000 &1.000& 1.000 &1.000 &1.000&1.000&1.000& 1.000& 1.000 &1.000  \\[1ex]
        & 100 & 1.000 & 1.000& 0.999 & 0.999& 1.000& 1.000& 1.000&1.000& 1.000&1.000& 1.000 &1.000 &1.000 \\[1.5ex]
\hline \\
$EV(1)$ & 25  & 1.000 &1.000 & 0.998&  0.999 &0.999& 0.999& 0.999&1.000& 0.997&0.999 &0.999 &1.000 &0.999   \\[1ex]
        & 50  & 1.000 & 1.000& 0.997 & 1.000 &1.000 &1.000 &1.000 &1.000&1.000 &1.000&1.000& 1.000 &1.000  \\[1ex]
        & 75  & 1.000 & 1.000& 0.999 & 0.999& 1.000& 1.000& 1.000&1.000& 1.000&1.000& 1.000 &1.000 &1.000  \\[1ex]
        & 100 & 1.000  & 1.000 & 1.000 & 1.000 &1.000 &1.000& 1.000&1.000& 1.000&1.000& 1.000 &1.000 &1.000   \\[1.5ex]
\hline
\end{tabular}
\end{table}
\end{landscape}

\begin{small}
\begin{table}  
 \centering
 \caption{Empirical size and power for $\widehat{\Delta}_{I}$ at 0.05 level of significance}
 \vspace{0.2cm}
 \label{table:6} 
 \fontsize{9pt}{11pt}\selectfont
   \resizebox{13.5 cm}{!}{
    \begin{tabular}{ccccccccc}
    \hline\vspace{0.1cm}
    n &  P(5) & $\gamma(0.5)$  & LF(0.5)   & BE(0.5)   & TP(0.5)  & IB(0.5)  & B(0.5)  &  EV(0.5)   \\
\hline
\multicolumn{9}{c}{20\% censoring} \\ 
\hline
50  & 0.038 & 0.178 & 0.682 & 0.266 & 0.924 & 0.924 & 0.980 & 0.689  \\[1ex]
75  & 0.044 & 0.264 & 0.776 & 0.298 & 1.000 & 0.998 & 1.000 & 0.726     \\[1ex]
100 & 0.051 & 0.329 & 0.889 & 0.319 & 1.000 & 1.000 & 1.000 &  0.871  \\
\hline
\multicolumn{9}{c}{40\% censoring} \\ 
\hline
50  & 0.041 & 0.334& 0.756 & 0.239 & 0.998 & 0.948 & 1.000 & 0.776  \\[1ex]
75  & 0.049 & 0.390 & 0.789 & 0.262 & 1.000 & 1.000 & 1.000 &  0.810    \\[1ex]
100 & 0.047 & 0.400 & 0.830 & 0.346 & 1.000 & 1.000 &1.000  &  0.784  \\ 
\hline
 \end{tabular}%
   }
  \label{ }%
\end{table}%

\begin{table}  
 \centering
 \caption{Empirical size and power for $\widehat{\Delta}_{M}$ at 0.05 level of significance}
 \vspace{0.2cm}
 \label{table:7} 
 \fontsize{9pt}{11pt}\selectfont
   \resizebox{13.5 cm}{!}{
    \begin{tabular}{ccccccccc}  
    \hline\vspace{0.1cm}
  n &P(5)& $\gamma(0.5)$& LF(0.5)& BE(0.5)& TP(0.5)& IB(0.5)& B(0.5)&EV(0.5) \\[1ex]
\hline
\multicolumn{9}{c}{20\% censoring} \\ 
\hline
50  & 0.043 & 0.224 & 0.644 & 0.219 & 0.870 & 0.982 & 1.000 & 0.733 \\[1ex]
75  & 0.049 & 0.252 & 0.793 & 0.366 & 0.995 & 1.000 & 1.000 & 0.842 \\[1ex]
100 & 0.053 & 0.381 & 0.841 & 0.403 & 1.000 & 1.000 & 1.000 & 0.905   \\ 
\hline
\multicolumn{9}{c}{40\% censoring} \\
\hline
50  & 0.038 & 0.236 & 0.598 & 0.267 & 0.880 & 1.000 & 1.000 & 0.793  \\[1ex]
75  & 0.049 & 0.307 & 0.768 & 0.354 & 1.000 & 1.000 & 1.000 & 0.890  \\[1ex]
100 & 0.055 & 0.419 & 0.839 & 0.389 & 1.000 & 1.000 & 1.000 & 0.934 \\
\hline
 \end{tabular}%
   }
  \label{ }%
\end{table}%
\end{small}

\section{Data analysis}\label{sec6}

This section utilizes R software to perform numerical simulations to investigate the proposed test procedure. 

\subsection{Complete case}\label{sec6.1}
Two real data sets are examined to determine whether the observed data are consistent with the theory that they follow a Pareto type-I distribution.

\subsection*{Illustration 1:}
The exceedances of flood maxima from the Wheaton River, located near Carcross in Canada's Yukon Territory (\cite{choulakian2001goodness}), were examined. Table \ref{dataset1} shows the dataset used for this research, which contains $72$ exceedance readings for the years $1958$ and $1984$. All values have been rounded to the nearest tenth of a cubic meter per second $(m^3/s)$. Using these data, the goodness of fit of the transmuted Pareto distribution versus the simple Pareto distribution was tested within a Bayesian framework by \cite{aslam2020bayesian}. 

\begin{table}[h!] 
\centering
\caption{Exceedances of 
Wheaton River flood data} \label{dataset1}
\vspace{0.2cm}
\fontsize{9pt}{11pt}\selectfont
\resizebox{13.5 cm}{!}{
\begin{tabular}{ccccccccccccccccccccccccccccccccccc } 
\hline 
    1.7 &2.2& 14.4& 1.1& 0.4& 20.6& 5.3& 0.7& 13.0& 12.0& 9.3& 1.4& 18.7& 8.5& 25.5\\[1ex]
    11.6& 14.1& 22.1& 1.1& 2.5& 14.4& 1.7& 37.6& 0.6&2.2& 39.0& 0.3& 15.0& 11.0&7.3\\[1ex]
    22.9& 1.7& 0.1& 1.1& 0.6& 9.0& 1.7& 7.0& 20.1& 0.4&14.1& 9.9& 10.4& 10.7& 30.0 \\[1ex]
    3.6& 5.6& 30.8& 13.3& 4.2& 25.5& 3.4& 11.9& 21.5& 27.6& 36.4&2.7& 64.0& 1.5& 2.5\\[1ex]  
    27.4& 1.0&27.1& 20.2 &16.8& 5.3& 9.7& 27.5& 2.5& 27.0& 1.9& 2.8 \\
\hline
\end{tabular}
}
\end{table}

\subsection*{Illustration 2:}
This dataset comprises the financial costs associated with wind-related disasters in $40$ different incidents during $1977$, rounded to the nearest million US dollars. The rounding of values creates an erroneous clustering effect that may complicate the determination of whether the data fits the Pareto distribution. The de-grouping procedure and the evaluation of the Pareto distribution for the dataset displayed in Table \ref{dataset2} are discussed in detail by \cite{ndwandwe2023testing} under various parameter estimation setups.
\begin{table}[h!] 
\centering
\caption{Wind catastrophes
de-grouped data set} \label{dataset2}
\vspace{0.2cm}
\fontsize{9pt}{11pt}\selectfont
\resizebox{13.5 cm}{!}{
\begin{tabular}{ccccccccccccccccccccccccc } 
\hline
     1.58& 1.65& 1.73& 1.81& 1.88& 1.96& 2.04& 2.12& 2.19& 2.27& 2.35& 2.42\\[1ex]
     2.70& 2.90& 3.10& 3.30& 3.75& 4.00& 4.25& 4.70& 4.90& 5.10& 5.30& 5.70\\[1ex]
     5.90& 6.10& 6.30& 7.83& 8.17& 9.00& 15.00& 17.00& 22.00& 23.00& 23.83 &24.17 \\[1ex]
     25.00& 27.00& 32.00& 43.00 \\
\hline
\end{tabular}
}
\end{table}

The following bootstrap algorithm is applied to determine the critical points. The calculated values of $\widehat{\Delta}_{I}$ and $\widehat{\Delta}_{M}$ test statistics, along with the critical points and the moment-based estimator for these datasets, are reported in Tables \ref{table:result1}-\ref{table:result2}. At the $5\%$ significance level, it is demonstrated that the tests fail to reject the null hypothesis regarding the exceedances of the Wheaton River flood data and the wind catastrophe data.

\begin{table}[h!] 
\centering
\caption{Analysis of complete data sets}
\vspace{0.2cm}
\label{table:result1}
\fontsize{9pt}{11pt}\selectfont
\resizebox{10.5 cm}{!}{
\begin{tabular}{c c c c c} 
 \hline
 \multicolumn{5}{c}{ $\widehat{\Delta}_{I}$ } \\[0.5ex] 
 \hline
 & Test statistic & $C_1$ & $C_2$ & $\hat{\alpha}$  \\ [0.5ex] 
 \hline 
 Illustration 1 & -0.2074  & -0.9313 & 0.3042 & 1.0892 \\[0.5ex] 
 Illustration 2 &  0.4653 & 0.3588 & 0.5827 & 1.1215\\ [0.5ex] 
 \hline
 \end{tabular}}
\end{table}

\begin{table}[h!] 
\centering
\caption{Analysis of complete data sets}
\vspace{0.2cm}
\label{table:result2}
\fontsize{6pt}{8pt}\selectfont
\resizebox{10.5 cm}{!}{
\begin{tabular}{c c c c c} 
 \hline
 \multicolumn{4}{c}{ $\widehat{\Delta}_{M}$ } \\[0.5ex] 
 \hline
  & Test statistic & $C_3$   & $\hat{\alpha}$  \\ [0.5ex] 
 \hline 
 Illustration 1 & 0.0108   &  0.4050   & 1.0892  \\[0.5ex] 
 Illustration 2 & 0.2223 &  0.3099   & 1.1215 \\ [0.5ex] 
 \hline
\end{tabular}}
\end{table}

\subsection{ Censored case}\label{sec6.2}
Using the re-weighting methods described in Section \ref{sec3.2} and \ref{sec4.2}, the asymptotic null variances of $\widehat{\Delta}_{I}$ and $\widehat{\Delta}_{M}$ are calculated. The data on the failure lifetime of items, available in \cite{saldana2010goodness}, is analyzed to test the Pareto distribution assumption. The dataset consists of $30$ items, with $10$ of them being censored, resulting in $33.33\%$ of censored observations. The corresponding moment-based estimator is $\hat{\alpha} = 1.0319$. The test statistics, calculated as $\widehat{\Delta}_{I} = 0.5150$ and $\widehat{\Delta}_{M} = -19.8816$, indicate that the null hypothesis that the data follow a Pareto type I distribution is not rejected at the $5\%$ significance level.

\section{Summary}\label{sec7}
In this paper, two tests based on Stein's type identity are proposed for testing the Pareto type-I distribution for complete data. The modification of our tests to incorporate the censored observations is discussed. The asymptotic distributions of the proposed test statistics are obtained for both cases based on U-statistic theory. A simulation study has been carried out to assess the performance of the proposed test procedures. Notably, across all provided alternatives and for both sample sizes, greater powers are exhibited by our tests utilizing the statistics $\Delta_{I}$ and $\Delta_{M}$ compared to the existing tests. Finally, the proposed methodologies are implemented on various compelling real-world data scenarios, including the exceeds of Wheaton River flood and wind catastrophe data sets. It is revealed that both datasets suggest that the Pareto type-I distribution can be adopted as a reasonably good model.
\medskip

 \bibliographystyle{apalike}
 \bibliography{02pareto}

\end{document}